\documentclass{iopart}
\usepackage{iopams}

\newcommand{\Sc}{Schr\"odinger equation }
 \def\be{\begin{equation}}
\def\ee{\end{equation}}
\def\bea{\begin{eqnarray}}
\def\eea{\end{eqnarray}}
\def\vfi{\varphi}
\newcommand{\cU}{{\cal U}}
\newcommand{\cV}{{\cal V}}

\newtheorem{Th}{Theorem}
\newtheorem{lemma}{Lemma}

\def\baselinestretch{1.5}

\begin{document}

 \large

\title{Chains of Darboux transformations for the matrix \Sc }

\author{
 Boris F Samsonov$^{\dag}$\footnote[3]{On leave from
 Physics Department of Tomsk State
 University, 634050 Tomsk, Russia.
 } and
A~A~Pecheritsin$^{\P}$ }

\address{\dag\ Departamento de F\'{\i}sica Te\'orica, Universidad de
Valladolid,  47005 Valladolid, Spain}
\address{$^{\P}$Physics Department, Tomsk State
 University, 634050 Tomsk, Russia.}

\ead{\mailto{samsonov@phys.tsu.ru}, \mailto{pecher@ido.tsu.ru}
 }

\begin{abstract}
\noindent
 Chains of Darboux transformations for the matrix Schr\"odinger equation are
 considered. Matrix generalization of the well-known for the scalar equation Crum-Krein
 formulas for the resulting action of such chains is given.
\end{abstract}


\medskip
\medskip


\medskip
\medskip

\textbf{Corresponding Author}:

B F Samsonov

Departamento de F\'\i sica Te\'orica

Universidad de Valladolid

47005 Valladolid, Spain

\medskip

E-mail: {\it boris@metodos.fam.cie.uva.es\/}

\newpage


\section{Introduction}

Let us consider the matrix \Sc
 \be \label{1}
 h_0\Psi_E = E\Psi_E \,,\quad
  h_0=-D^2+V_0(x)\,,\quad D\equiv \frac{d}{dx}
 \ee
 where $V_0(x)$ is an $n\times n$ Hermitian matrix with $x$-dependent entries,
 $\Psi_E=(\psi_{1E},\ldots ,\psi_{nE})^t$ is a vector
 of an $n$-dimensional linear space,  and $E$ is a number which
 plays an essential role in different physical
 applications.
 For instance, a  multichannel quantum system is described by
 this equations \cite{ZC}.
 One of the most interesting applications of this equation consists
 in
 the possibility to involve the supersymmetric quantum mechanics
\cite{ACDIJMP}
 for describing
 scattering of composite particles like atom-atom  or
 nucleon-nucleon collisions \cite{SBPRC61,SBPRC62,SBPRL}.
 In particular, in this way one was
 able to interpret an ambiguity between shallow and deep
 potentials of  the nucleon-nucleon interaction
 \cite{SBPRL}.
To get a qualitative result
 authors of Refs \cite{SBPRC61,SBPRC62,SBPRL}
  apply  supersymmetric transformations successively
 which require
 performing a lot of additional work. In this way one is able to
 realize  only few transformation steps. We believe that
 the progress in applications of this method is essentially delayed
 because of the absence of a simple possibility to get rid of
 intermediate Hamiltonians and go directly to the final
 result of a chain of transformations.
 We notice  that such a
 possibility exists for the usual (scalar) \Sc
 being given by the known Crum-Krein determinant formulas
\cite{Kr,Krein}
 which made it
 possible to get a number of new interesting applications of
 supersymmetric quantum mechanics in describing
 the nucleon-nucleon scattering  \cite{SSPRC}.
 Nevertheless, this problem has been tackled by Goncharenko
 and Veselov \cite{GV} when they realized that Gelfand-Retakh
 quasideterminants \cite{GR} may be used for this purpose.
 We want to point out that although their method gives a
 solution in principle  this is is very complicated and difficult for practical
 realization since it involves a matrix calculus with matrices
 defined over a noncommutative ring and in particular it is
 necessary to
 invert  such matrices.

 In this paper we prove alternative formulas where only usual
 determinants are involved. They are very similar to the known
 Crum-Krein determinant formulas and can be considered as their
 straightforward generalizations.


\section{First order transformation}

 We follow the definition of the Darboux transformation operator given
 by Goncharenko and Veselov \cite{GV}
  defining  it as a first order
 differential operator with matrix-valued coefficients
\be \label{Li}
 L=L_0(x)+L_1(x)D
 \ee
 intertwining $h_0$ and $h_1$
\be \label{Irel}
 Lh_0=h_1L
 \ee
 where $h_0$ is introduced above and $h_1$ is defined by the potential
 $V_1$
 \be
\quad h_1=-D^2+V_1(x)
 \ee
  If such un operator is found one can get eigenfunctions of $h_1$,
 by the simple action of $L$ on the  eigenfunctions of $h_0$,
 \be\label{2}
 \Phi_E =L\Psi_E=(\vfi_{1E},\ldots ,\vfi_{nE})^t\,,
 \quad
 h_1\Phi_E = E\Phi_E
 \ee

Since the equations (\ref{1}) and (\ref{2}) are homogenous,
 without loss of generality  we can
 put $L_1$ equal to the identity, $L_1=1$
 (we suppose ${\rm det}L_1\ne 0$).
 After inserting $L$ into the
intertwining relation (\ref{Irel}) we get a  system of equations
for $L_0$ and the transformed potential $V_1$. It is not difficult
to find the solution to this system \cite{GV}. Thus, $L$ is given
by
 \be
 \label{L}
 L=D- F\,,\quad F= \cU '\cU^{-1}
 \ee
and for the potential $V_1$ one obtains
 \be
 V_1=V_0-2F'
 \ee
The matrix-valued function $\cU$ is a solution to the equation
\be\label{1U}
h_0\cU = \cU \Lambda
\ee
where $\Lambda$ is a constant matrix.

 The known supersymmetric approach \cite{ZC}-\cite{SBPRL} is based on the factorization of the
 Hamiltonian
 \be  \label{fact}
 h_0=L^+L+\lambda I \,,\quad \lambda \in \Bbb R
 \ee
  where $L^+$ is defined
 with the help of the formal relations: $D^+=-D$, $i^+=-i$,
 $(AB)^+=B^+A^+$ and $I$ is the identity matrix.
 To compare our method with this technique let us
 consider the superposition of $L$ and its conjugate.
 After a simple algebra one finds
 \be
 L^+L=h_0-\cU\Lambda\cU^{-1}
 \ee
In particular, when $\Lambda =\lambda I$,
 we recover the factorization  (\ref{fact})
 giving rise to the supersymmetry
 with $\lambda $ meaning the factorization constant.
Similarly, the inverse superposition gives
\be
LL^+=h_1-\cU\Lambda\cU^{-1}
\ee

If $\Lambda$ is a diagonal matrix
 $\lambda = \mbox{diag}(\lambda_1,\ldots \lambda _N)$ then the
 system of equations (\ref{1U}) just takes  the form of the
 \Sc for the columns $U_j=(u_{j,1},\ldots , u_{j,n})^t$ of the matrix
 $\cU=(U_{1},\ldots ,U_n)$
 \be
 h_0U_{j}=\lambda_jU_{j}\,,\qquad j=1,\ldots ,n
 \ee
  This means that if we know solutions of the \Sc (\ref{1}) then solutions of
 Eq. (\ref{1U}) are also known for the diagonal form of the
 eigenvalue matrix $\Lambda$. Therefore in what follows we will
 consider only diagonal $\Lambda$'s.

\section{Chains of Darboux transformations}

 Now we want to consider chains of transformations defined in the previous section.
 Chains
 appear naturally if we notice that  if    sufficiently many matrix solutions
 to the initial equation are known then any such a
 solution is transformed into a matrix solution of the new equation and,
 hence, the latter may play the role of the initial equation for the next
 transformation step etc.

 Suppose we know $N$ matrix solutions of Eq. (\ref{1U})
 corresponding to different eigenvalue matrices $\Lambda_k\ne
 \Lambda_l$
 \be
 h_0\cU_k=\cU_k\Lambda _k\,,\quad k=1,\ldots ,N
 \ee
 For the first transformation step we
 take the function $\cU_1$ and according to (\ref{L}) construct the transformation
 operator
 \begin{equation} \label{L1}
 L_{1 \leftarrow 0} =D  - \cU_1'\cU_1^{-1}
 \end{equation}
 We notice that it can be applied not only on vector-valued
 functions like $\Psi_E$ but also on matrix-valued like $\cU_{\,2}$,
 \ldots , $\cU_N$.
 In this way we get the matrix solutions
 $\cV_2=  L_{1 \leftarrow 0}\cU_{\,2}$, \ldots , $\cV_N=  L_{1 \leftarrow 0}\cU_N$
 of  the equation with the potential
\be
 V_1=V_0-2F_1'\,,\quad F_1=\cU_1'\cU_1^{-1}
\ee
 Now $\cV_2$ can be taken
 as transformation function for the Hamiltonian $h_1=-D^2+V_1$
 to produce the potential
 \be
  V_2=V_1-2(\cV_2'\cV_2^{-1})'=V_0-2F_2'\,,\quad
  F_2=F_1+\cV_2'\cV_2^{-1}
  \ee
 and the transformation operator
 $L_{2 \leftarrow 1} =D  - \cV_2'\cV_2^{-1}$
 and so on, till one gets the potential
 \be \label{ReVN}
 V_N=V_0-2F_N'
  \ee
 with $F_N$ defined recursively
 \be \label{Re1}
  F_N=F_{N-1}+2Y_N'Y_N^{-1}\,,\quad N=1,2,\ldots \ \,,\quad F_0=0
 \ee
 and $Y_N$ being the matrix-valued solution at $(N-1)$th step of
 transformations
 \be  \label{Re2}
 Y_N = L_{(N-1) \leftarrow (N-2)}\ldots L_{2 \leftarrow 1}L_{1 \leftarrow 0} \cU_N
 \equiv  L_{(N-1) \leftarrow 0} \cU_N
 \ee
 which produces  the final transformation
 operator $L_{N \leftarrow (N-1)}=-D+Y_N'Y_N^{-1}$.

 To get in this way the final potential $V_N$ one has to
 calculate all intermediate transformation functions performing
 a huge amount of the numerical work even for the scalar case.
 In practical calculations one is able to perform only few steps
 which restricts considerably possible applications of the method.
 Fortunately, for the scalar case there exists what that are
 called Crum \cite{Kr} or Crum-Krein \cite{Krein} formulas which allow to omit
 all intermediate steps and go directly from $h_0$ to $h_N$.
 The function
 \be \label{fiNscal}
 \varphi_E=\frac{|W(u_1,\ldots ,u_N,\psi_E )|}{|W(u_1,\ldots ,u_N)|}
 \ee
 is an eigenfunction of the Hamiltonian $h_N$ with the potential
 \be   \label{VNscal}
 V_N=V_0-2 \left(
 \frac{|\widetilde W(u_1,\ldots ,u_N)|}{|W(u_1,\ldots  ,u_N)|}\right)'
 \ee
 provided all $u_k$, $k=1,\ldots ,N$ and $\psi_E$ are eigenfunctions of the initial
 Hamiltonian $h_0$ with the scalar potential $V_0$: $h_0=-D^2+V_0$,
 $h_0u_k=\alpha_ku_k$, $h_0\psi_E=E\psi_E$.
 Here and in what follows the symbol $|\cdot |$ means the usual determinant,
  $W(u_1,\ldots ,u_N)$ is the Wronsky matrix
\be
 W(u_1,\ldots ,u_N)=
 \left(
   \begin{array}{cccc}
   u_1 & u_2 & \ldots & u_N \\
   u_1' & u_2' & \ldots & u_N' \\
   \ldots  &  \ldots &  \ldots &  \ldots \\
  u_1^{(N-1)} & u_2^{(N-1)} & \ldots & u_N^{(N-1)}
  \end{array}
   \right)
 \ee
  and
 the matrix $\widetilde W(u_1,\ldots ,u_N)$ is obtained from
 $W(u_1,\ldots ,u_N)$
  by replacing its last row composed of $u_k^{(N-1)}$ with
  $u_k^{(N)}$, $k=1,\ldots ,N$.
  Of course the determinant $|\widetilde W(u_1,\ldots ,u_N)|$
   is nothing but the derivative of the determinant of the Wronsky
   matrix $|W(u_1,\ldots ,u_N)|$ but we write the second logarithmic
   derivative of the Wronskian $|W(u_1,\ldots ,u_N)|$ in
   (\ref{VNscal}) as the first derivative of the ratio of
   corresponding determinants to stress the similarity of this
   scalar formula and its matrix generalization below.
  The formula (\ref{fiNscal}) really defines for the scalar case the
  superposition of the operators of the type (\ref{L1})
  with the replacement of matrix-valued functions $\cU_k$ by usual
  functions $u_k$
  \be \label{L0N}
 L_{N \leftarrow 0}=
 L_{N \leftarrow (N-1)}\ldots L_{2 \leftarrow 1}L_{1 \leftarrow 0}
  \ee
  \be
 \varphi_E = L_{N \leftarrow 0}\psi_E
   \ee

 We shall prove below generalizations of the
 formulas (\ref{fiNscal}) and (\ref{VNscal}) to the matrix case
 meaning that we shall solve the recursion defined by (\ref{Re1}),
 (\ref{Re2}) and find the superposition of the operators (\ref{L0N})
 but first we need to introduce some new notations and to prove
  an additional statement.

\subsection{Notations}

Define first the   $nN$-dimensional square  Wronsky matrix
\begin{equation}  \label{W-def}
  W(\cU_1,\ldots, \cU_N)=
  \left(
  \begin{array}{cccc}
  \cU_1 & \cU_2 & \ldots & \cU_N \\
  \cU_1' & \cU_2' & \ldots & \cU_N' \\
  \ldots  &  \ldots &  \ldots &  \ldots \\
 \cU_1^{(N-1)} & \cU_2^{(N-1)} & \ldots & \cU_N^{(N-1)}
 \end{array}
  \right)
  \end{equation}
 Here $\cU_k$ are $n \times n$ matrices
\be
 \cU_k=  \left(
  \begin{array}{cccc}
  u_{1,1;k} & u_{1,2;k} & \ldots & u_{1,n;k} \\
  u_{2,1;k} & u_{2,2;k} & \ldots & u_{2,n;k} \\
  \ldots  &  \ldots &  \ldots &  \ldots \\
 u_{n,1;k} & u_{n,2;k} & \ldots & u_{n,n;k}
 \end{array}
  \right) \,,\quad k=1,\ldots ,N
 \ee
 with columns
 being $n$-dimensional vectors $U_{j;k}=(u_{1,j;k}, \ldots
 ,u_{n,j;k})^t$
 so that
$\cU_k=(U_{1;k},\ldots ,U_{n;k})$, $k=1,\ldots ,N$.
 We can also present $\cU_k$ as a collection of rows
 $U_k^j=(u_{j,1;k},\ldots u_{j,n;k})$, $j=1,\ldots ,n$,
 $\cU_k=(U_k^1,\ldots ,U_k^n)^t$.
  It will be convenient to present (\ref{W-def}) also in the form
 \begin{equation}\label{WNnew}
 W(\cU_1,\ldots, \cU_N)=
 \left(
 \begin{array}{cc}
  W(\cU_1, \ldots, \cU_{N-1}) &
    \begin{array}{c}
  \cU_{N} \\  \cU_{N}' \\ \ldots
    \end{array}
  \\
       \begin{array}{rrr}
 \cU_1^{(N-1)} &   \ldots &  \cU_{N-1}^{(N-1)}
       \end{array}
     &
   \cU_{N}^{(N-1)}
 \end{array}
  \right)
  \end{equation}
 stressing its recursion nature.

 Introduce also the matrix
  \begin{equation}\label{Wj}
  W_{jE}(\cU_1, \ldots, \cU_{N})=
  \left(
  \begin{array}{cc}
   W(\cU_1, \ldots, \cU_{N}) &
     \begin{array}{c}
   \Psi_E \\ \Psi'_E \\ \ldots \\ \Psi_E^{(N-1)}
     \end{array}
   \\
        \begin{array}{rrr}
  (U_1^{j})^{(N)} &   \ldots &  (U_N^{j})^{(N)}
        \end{array}
      &
    \psi_{jE}^{(N)}
  \end{array}
   \right)
    \end{equation}
 recalling that $U_k^{j}$ to be the $j$th row of the matrix $\cU_k$.

  We shall also need  the following  matrix
\begin{equation}\label{Wjk}
 W_j^i(\cU_1, \ldots, \cU_{N-1})=
 \left(
 \begin{array}{cc}
  W(\cU_1, \ldots, \cU_{N-1}) &
    \begin{array}{c}
  U_{i;N} \\  U_{i;N}' \\ \ldots \\  U_{i;N}^{(N-2)}
    \end{array}
  \\[2.5em]
       \begin{array}{rrr}
 (U_1^{j})^{(N-1)}  & \ldots &
 (U_{N-1}^{j})^{(N-1)}
       \end{array}
     &
   u_{j,i;N}^{(N-1)}
 \end{array}
  \right)
  \end{equation}
 First we notice that this is the previous matrix where $N$
 is replaced with $N-1$ and $\Psi_E$ with the vector $U_{i;N}$.
 Another useful remark is that the determinants
 $|W_j^i(\cU_1, \ldots, \cU_{N-1})|$,
 $i,j=1,\ldots ,n$ are nothing
 but minors embordering the block $W(\cU_1, \ldots, \cU_{N-1})$ in
 the determinant of the matrix (\ref{WNnew}). (For the definition of embordering minors
 see Appendix.)

Finally we   introduce the matrices
 $W_{i,j}(\cU_1, \ldots, \cU_N)$, $i,j=1,\ldots ,n$,
 constructed from the
 Wronsky matrix (\ref{W-def}) where the last matrix row
 composed of matrices $\cU_k^{(N-1)}$ is replaced by
 $\cU_k^{ij}$, $k=1,\ldots ,N$:
  \begin{equation}\label{Wij}
W_{i,j}(\cU_1, \ldots, \cU_N)=
  \left(
  \begin{array}{cccc}
  \cU_1 & \cU_2 & \ldots & \cU_N \\
  \cU_1' & \cU_2' & \ldots & \cU_N' \\
  \ldots  &  \ldots &  \ldots &  \ldots \\
 \cU_1^{(N-2)} & \cU_2^{(N-2)} & \ldots & \cU_N^{(N-2)}\\
 \cU_1^{ij} & \cU_2^{ij} & \ldots & \cU_N^{ij}
 \end{array}
  \right)
  \end{equation}
 The matrices $\cU_k^{ij}$, $i,j=1,\ldots ,n$ are constructed from
 the matrix
 $\cU_k^{(N-1)}$ by replacing its $j$th row with the $i$th row of
 the matrix  $\cU_k^{(N)}$.

 \subsection{Main Lemma}

 In this subsection we prove a lemma we are using in the
 proof of  theorems below.
 Moreover, in proving it as well as the theorems  we are
 using the {\it Sylvester identity} \cite{Gantmaher} which is
 formulated in the Appendix.

 Consider the matrix
 \begin{equation}
 A=\left(
 \begin{array}{cccccc}
 a_{1,1} & \ldots & a_{1,p} & a_{1,p+1} & \ldots & a_{1,p+n}\\
        &  \ldots &       &          & \ldots &         \\
 a_{p,1} & \ldots & a_{p,p} & a_{p,p+1} & \ldots & a_{p,p+n}\\
 b_{1,1} & \ldots & b_{1,p} & b_{1,p+1} & \ldots & b_{1,p+n}\\
 b_{2,1} & \ldots & b_{2,p} & b_{2,p+1} & \ldots & b_{2,p+n}
 \end{array}
 \right)
\end{equation}
 Let $a$ be the $p\times p$ submatrix of $A$
 with the entries $a_{i,j}$, $i,j=1,\ldots ,p$.
 Denote $m_{jk}$ the minor of $A$ embordering $a$ with
 $j$th ($j=1,2$) row composed of  $b_{j,i}$, $j=1,2$,
 $i=1,\ldots ,p$ and $k$th column ($p<k\le p+n$).
 Let also $m_{jk}^{ts}$ be the minor obtained from  $m_{jk}$
 by replacing its $s$th row composed of
  $a_{s,j}$, ($s\le p$) with $(p+t)$th row composed of
   $b_{t,i}$ ($t=1,2$).
   Let now $a^{ts}$ be obtained from $a$ with the help of the same
   replacement, i.e. with the replacement of its $s$th row
   composed of $a_{s,j}$, ($s\le p$) by $(p+t)$th row of $A$ composed of
    $b_{t,j}$ ($t=1,2$).

  \begin{lemma}\label{lm:minors}
 If  $|a|\ne 0$ then
 the following determinant identity takes place
\begin{equation}
 |a|\, m_{jk}^{ts} = |a^{ts}| m_{jk} - |a^{js}| m_{tk}
 \label{m_jk^ts}
\end{equation}
\end{lemma}

\noindent
{\it Proof}.
 Consider an auxiliary square matrix
 \begin{equation}
 A_{jt}=\left(
 \begin{array}{ccccc}
 a_{1,1} & \ldots & a_{1,p} & a_{1,k} & 0 \\
        &  \ldots &       &        &  \ldots\\
 a_{s,1} & \ldots & a_{s,p} & a_{s,k} & 0 \\
        &  \ldots &       &        & \ldots \\
 a_{p,1} & \ldots & a_{p,p} & a_{p,k} & 0 \\
 b_{j,1} & \ldots & b_{j,p} & b_{j,k} & 1 \\
 b_{t,1} & \ldots & b_{t,p} & b_{t,k} & 1
 \end{array}
 \right).
\end{equation}
where $j,t=1,2$ and the last column containing only two nonzero
entries. Take its main minor  $|a|$.
 There are only four minors of $A_{jt}$ embordering $|a|$. Minors
  $m_{jk}$ and $|a|$ emborder it by the row $b_{j,i}$,  $i=1,\ldots ,p$
 and the next to last and the last columns respectively and minors
  $m_{tk}$ and  $|a|$ emborder it with the last row and the same
  columns.
 According to the Sylvester identity the determinant composed of
 these embordering minors is equal
 \begin{equation}
m_{jk} |a| - m_{tk} |a| = |a| |A_{jt}|. \label{detA_jt}
\end{equation}
where we can cancel $|a|$ since it is supposed to be different
from zero.

Now interchange in the matrix $A_{jt}$  the $s$th and the last
rows to get
\begin{equation}
 \widetilde{A}_{jt}=\left(
 \begin{array}{ccccc}
 a_{11} & \ldots & a_{1p} & a_{1k} & 0 \\
        &  \ldots &       &        &  \ldots  \\
 b_{t1} & \ldots & b_{tp} & b_{tk} & 1 \\
        &  \ldots &       &        &  \ldots \\
 a_{p1} & \ldots & a_{pp} & a_{pk} & 0 \\
 b_{j1} & \ldots & b_{jp} & b_{jk} & 1 \\
 a_{s1} & \ldots & a_{sp} & a_{sk} & 0
 \end{array}
 \right)
 \hspace{-.5em}
 \begin{array}{l}
 \vphantom{0}\\
  \vphantom{0}\\
  \leftarrow {\rm s}t\!h \ {\rm row}\\
   \vphantom{0}\\
    \vphantom{0}\\
     \vphantom{0}\\
      \vphantom{0}\\
 \end{array}
\end{equation}
 The upper-left block of this matrix of dimension $p \times p$ is
 the above introduced matrix  $a^{ts}$. Let us find embordering
 minors for this submatrix.
 It is clear that  $m_{jk}^{ts}$  and $-m_{tk}$ emborder it with
 the next to last column and the the next to last and the last
 rows respectively. The minor embordering   $a^{ts}$ with the next
 to last row and the last column has in the last column only two
 nonzero entries which are equal to one. Therefore we can
 decompose it on this column and after corresponding interchange
 of the rows one gets for it the expression  $|a^{ts}|-|a^{js}|$.
 The last embordering minor is equal to  $-|a|$ which becomes
 evident after corresponding interchange of the rows.
 Once again we consider the determinant composed of these minors and
 calculate it using the Sylvester identity
 \begin{equation} \label{det_wA_jt}
 |\widetilde{A}_{jt}|\, |a^{ts}| = m_{tk}|a^{ts}| -  m_{tk}|a^{js}|  -|a| m_{jk}^{ts}
 \end{equation}
  Since $|\widetilde{A}_{jt}| = - |A_{jt}|$ the lemma follows
  from the equations (\ref{detA_jt}) and (\ref{det_wA_jt}).
  \hfill $\Box$

 \subsection{Transformation of vectors}

  In this Subsection we formulate and prove the theorem about the
  transformation of a vector $\Psi_E$ by a chain of
  transformations
  introduced at the beginning of this Section.

 \begin{Th}
 The resulting action of a chain of Darboux transformations
 applied to a vector
 \be
 \Psi_E=(\psi_{1E},\ldots ,\psi_{NE})^t
 \ee
 is the vector
 \be
 \Phi_E= L_{N \leftarrow (N-1)} \ldots L_{2 \leftarrow 1} L_{1 \leftarrow 0}\Psi_E =
 (\vfi_{1E},\ldots ,\vfi_{NE})^t
 \ee
 with the entries $\vfi_{jE}$ given by
 \begin{equation}\label{phi_j}
  \vfi_{jE}=\frac{|W_{jE}(\cU_1, \ldots, \cU_{N})|}{|W(\cU_1, \ldots, \cU_{N})|}
 \end{equation}
 where $W_{jE}(\cU_1, \ldots, \cU_{N})$ is given in $(\ref{Wj})$ and
 $W(\cU_1, \ldots, \cU_{N})$ is defined by $(\ref{W-def})$.
  \end{Th}

\noindent
{\it Proof}.
 To prove Theorem 1 we use the perfect induction method.
 So, first  we shall  prove it  for $N=1$.
 In this case
  \begin{equation}\label{LN1N}
\Phi_E = L_{1 \leftarrow 0}\Psi_E =(D  - \cU_1'\cU_1^{-1})\Psi_E =
(\vfi_{1E},\ldots ,\vfi_{nE})^t
  \end{equation}
 Denote $A_{ij}$ the cofactor of the element $u_{j,i;1}$ in the
 matrix $\cU_1$. Then according to the definition of the inverse
 matrix one has
 \begin{equation}
 (\cU^{-1} \Psi_E )_j=\frac{1}{|\cU _1|}\sum\limits_{i=1}^n
 A_{ij}\psi_{iE}
 \end{equation}
  which for the elements of the vector $\Phi_E$ implies
 \begin{equation} \label{fi_l}
  \varphi_{lE}=\partial \psi_{lE}- \frac{1}{|\cU_1 |}
  \sum\limits_{i,j=1}^n u_{l,j;1}' A_{ij}\psi_{iE} \equiv
  \frac{\Delta_{lE}}{|\cU_1 |}
 \end{equation}
 Consider now the matrix
 \begin{equation}  \label{W_j}
W_{jE}(\cU_1) = \left( \begin{array}{cc} \cU_1 & \begin{array}{r} \psi_{1E} \\
\ldots \\ \psi_{nE}
          \end{array} \\
\begin{array}{ccc}  u_{j,1;1}' & \ldots &  u_{j,n;1}'  \end{array}
 &  \psi_{jE}'
\end{array} \right)
\end{equation}
If we decompose the determinant $|W_{jE}(\cU_1)|$ on the elements
of the last row and all determinants appearing in this
decomposition except for the one coinciding with $|\cU_1 |$
decompose on the
 elements of the last column, the resulting expression will
   coincide exactly with the numerator of the right-hand side of (\ref{fi_l})
   meaning that
   \begin{equation}\label{L_Psi}
\varphi_{jE} = \frac{|W_{jE}(\cU_1)|}{|\cU_1|}
\end{equation}
  which proves
  the assertion for $N=1$.

 Suppose Theorem 1 to be true for the chain of $N-1$
 transformations meaning the following vector gives the resulting action
 of this chain:
 \begin{equation} \label{L_N-1_Psi}
{\Theta_E} = L_{(N-1) \leftarrow 0} \Psi_E = (\theta_{1E},\ldots ,
\theta_{nE})^t
\end{equation}
 where
 \begin{equation}\label{theta_j}
  \theta_{jE}=\frac{|W_{jE}(\cU_1, \ldots, \cU_{N-1})|}{|W(\cU_1, \ldots, \cU_{N-1})|}
 \end{equation}
 Now since
 $L_{N \leftarrow 0}= L_{N \leftarrow (N-1)}L_{(N-1) \leftarrow  0}$
 we have to apply the first order operator $L_{N \leftarrow (N-1)}$
 to the vector (\ref{L_N-1_Psi}) but first we need  to act with
 $L_{(N-1) \leftarrow  0}$ on the vectors $U_{i;N}$,
 $i=1,\ldots  ,n$ which form the columns of the matrix-valued
 transformation function $\cU_N$ to find the transformation
 function $Y_N =  L_{(N-1) \leftarrow 0} \cU_N $
  for the $N$th transformation step and determine the operator
  $L_{N \leftarrow (N-1)}=-D+Y'_NY_N^{-1}$.
 By supposition of (\ref{L_N-1_Psi}), (\ref{theta_j}) this result is given by
 \begin{equation}  \label{L_N-1_u_kN}
L_{(N-1) \leftarrow 0} U_{i;N} =(y_{1,i},\ldots ,y_{n,i})^t
\end{equation}
\begin{equation}\label{y_jk}
y_{j,i}=\frac{|W^i_j(\cU_1, \ldots, \cU_{N-1})|}{|W(\cU_1, \ldots,
\cU_{N-1})|} \,,\quad i,j=1,\ldots ,n
\end{equation}
 This means that the matrix
$Y_N$
 has $ y_{j,i}$ (\ref{y_jk}) as entries.
 If we notice that they coincide
 up to the constant factor  $1/|W(\cU_1, \ldots, \cU_{N-1})|$
 with the minors embordering the block
  $  W(\cU_1, \ldots, \cU_{N-1})$ in the matrix (\ref{WNnew})
  we can apply the Sylvester identity to calculate
  the determinant $|Y_N|$ which gives
  \begin{equation}\label{Y_N}
 |Y_N|=\frac{  |W(\cU_1, \ldots, \cU_{N})|}{  |W(\cU_1, \ldots, \cU_{N-1})|}
\end{equation}
 We need this determinant since the action of the first order
 Darboux transformation operator on a vector is given by
 (\ref{LN1N}),
   (\ref{L_Psi}) and
  (\ref{W_j})
 which implies that the vector
 \be  \label{FiE}
 \Phi_E = [D - Y_N' Y_N^{-1}] \Theta_E
 \ee
  has the entries
 \be    \label{vfie}
 \vfi_{jE}=|Y_{NE}^j|/|Y_N|\,,\quad j=1,\ldots ,n
 \ee
  with
 \be  \label{Y_N^j}
 Y_{NE}^j = \left(
 \begin{array}{cc} Y_N &
     \begin{array}{r} \theta_{1E} \\
 \ldots \\ \theta_{nE}
     \end{array} \\
  \begin{array}{rcr}  y_{j,1}' & \ldots & y_{j,n}'
  \end{array}
 & \theta_{jE}'
 \end{array} \right)
 \ee
  To calculate the entries of this matrix we need to differentiate
  $y_{j,i}$ (\ref{y_jk}) which yields
 \begin{equation}
  y_{j,i}'
 = -\frac{ |W(\cU_1, \ldots, \cU_{N-1})|'}
 {|W(\cU_1, \ldots, \cU_{N-1})|} y_{j,i} +
 \frac{ |W_j^i(\cU_1, \ldots, \cU_{N-1})|'}{|W(\cU_1, \ldots, \cU_{N-1})|}
 \label{dy_jk}
 \end{equation}
  Here we first calculate the derivative of the determinant
  $|W_j^i|$ keeping in this expression the derivative of its last row as a separate
  term
  \begin{equation}\label{dW_j^k}
   |W_j^i(\cU_1, \ldots, \cU_{N-1})|' = \Gamma_{ji}+\sum\limits_{m=1 (\neq j)}^n\Delta_{ji}^m
  \end{equation}
  Here $\Gamma_{ji}$ is the  determinant of the same matrix
   (\ref{Wjk}) where only the last
  row is differentiated and $\Delta_{ji}^m$
  is the determinant of the same matrix
  where in the last matrix row of the block
  $W(\cU_1, \ldots, \cU_{N-1})$ one has to differentiate
   only the $m$th row of the matrices
   $\cU_k^{(N-2)}$, $k=1,\ldots ,N$.

  We notice that the matrix of the determinant
  $\Delta_{ji}^m$ can also be obtained by interchanging one of the
  rows of matrices
   $\cU_k^{(N-2)}$
   with a row of matrices  $\cU_k^{(N-1)}$, $k=1,\ldots ,N$
   in the minor $|W_j^i(\cU_1, \ldots,\cU_{N-1})|$ when it is considered as a
    minor embordering the block
     $W(\cU_1, \ldots,\cU_{N-1})$ in the matrix
      (\ref{WNnew}).
   This matrix contains two rows with the derivatives of
   $(N-1)$th order of elements of matrices $\cU_k$
   in contrast with all minors embordering the block
    $  W(\cU_1, \ldots,\cU_{N-1})$  in the matrix (\ref{WNnew})
    which contain only one such a derivative.
  Therefore one is not able to apply the Sylvester identity
  to calculate this determinant. Just for this purpose we proved
  our main Lemma which implies
  \begin{equation} \label{WD}
  \begin{array}{l}
|W(\cU_1, \ldots,\cU_{N-1})|\Delta_{ji}^m \\[.5em]
=
 |W_{mm}(\cU_1, \ldots, \cU_{N-1})|\, |W_j^i(\cU_1, \ldots,
 \cU_{N-1})|\\[.5em]
  -
 |W_{jm}(\cU_1, \ldots, \cU_{N-1})|\, |W_m^i(\cU_1, \ldots, \cU_{N-1})|
 \end{array}
\end{equation}
 Now from (\ref{dW_j^k}) and (\ref{WD}) one gets
 \begin{eqnarray}
 &|W_j^i(\cU_1, \ldots, \hspace{-.7em} \cU_{N-1})|' =
  {\displaystyle\frac{1}{|W(\cU_1, \ldots,  \cU_{N-1})|}}
  \bigl\{\Gamma_{ji}|W(\cU_1, \ldots, \cU_{N-1})|
 & \nonumber \\
 &
 +
 \sum\limits_{m=1 (\neq j)}^n
\left[|W_{mm}(\cU_1, \ldots, \cU_{N-1})|\, |W_j^i(\cU_1, \ldots, \cU_{N-1})|\right.
&
 \\
&
 \left.
  - |W_{jm}(\cU_1, \ldots, \cU_{N-1})|\, |W_m^i(\cU_1, \ldots, \cU_{N-1})|\,\right] \bigr\}
  &\nonumber
\end{eqnarray}
 Inserting this into  (\ref{dy_jk}) and taking into account the
 relation
 \begin{equation}
 |W(\cU_1, \ldots, \cU_{N-1})|' = \sum\limits_{m=1}^n
 |W_{mm}(\cU_1, \ldots, \cU_{N-1})|
 \end{equation}
 which is a direct consequence of the structure of the matrices
 $W(\cU_1, \ldots, \cU_{N-1})$
 and $W_{mm}(\cU_1, \ldots, \cU_{N-1})$
 we get
 \begin{equation}    \label{dy_jk_fin}
  y_{j,i}'
 = \frac{1}{|W(\cU_1, \ldots, \cU_{N-1})|} \left[ \Gamma_{ji} -
 \sum\limits_{m=1}^n |W_{j,m}(\cU_1, \ldots, \cU_{N-1})| y_{m,i}
 \right]
  \end{equation}
 By the same means for the vector $\Theta_E$ (\ref{L_N-1_Psi}) we can get a similar
 relation.

 Thus, the determinant of the matrix  (\ref{Y_N^j}) can be written as a sum of
 two other determinants one of which has the last row to be a linear
 combination of other rows and, hence, this determinant vanishes.
 The matrix of another determinant consists of minors embordering
 block  $W(\cU_1, \ldots, \cU_{N-1})$ in the determinant  (\ref{Wj})
 (up to a common factor).
 Applying once again the Sylvester identity one finally gets
 \begin{equation}
 |Y_{NE}^j| = \frac{|W_{jE}(\cU_1, \ldots,  \cU_{N-1})|}%
 {|W(\cU_1, \ldots,  \cU_{N-1})|}
 \label{Y_N^k_fin}
 \end{equation}
  which together with (\ref{FiE}), (\ref{vfie}) and (\ref{Y_N}) proves the theorem.
  \hfill $\Box$

  \subsection{Transformation of potential}

 According to (\ref{ReVN})  to find the potential
 resulting from a chain of $N$ Darboux transformations we have to
 resolve the recursion defined in (\ref{Re1}) and (\ref{Re2}).
 This is done by the following
 \begin{Th}
 Let the matrix $F_N$ be defined by the recursion relation
  $F_N = F_{N-1} + Y_N' Y_N^{-1}$, $N=1,2,\ldots $
  with the initial condition $F_0 = 0$
  and   $Y_N =  L_{(N-1) \leftarrow 0} \cU_N$
 with the operator $ L_{(N-1) \leftarrow 0}=
 L_{(N-1) \leftarrow (N-2)}\cdot \ldots \cdot
 L_{2 \leftarrow 1} \cdot L_{1 \leftarrow 0}$
 defined in the Theorem 1.
 Then the elements $f^N_{i,j}$ of the matrix $F_N$
 are expressed in terms of transformation functions $\cU_k$,
 $k=1,\ldots ,N$
 as follows
 \begin{equation}\label{defV_N}
f^N_{i,j} = \frac{|W_{ij}(\cU_1, \ldots, \cU_N)|}{|W(\cU_1,
\ldots, \cU_N)|}
\end{equation}
 where $W(\cU_1, \ldots, \cU_N)$ is defined in $(\ref{W-def})$
 and $W_{ij}(\cU_1, \ldots, \cU_N)$ is given by $(\ref{Wij})$.
 \end{Th}

\noindent
{\it Proof}. This theorem is also proved by the perfect induction
 method. For $N=1$ one has $Y_1=\cU_1$ and $W(\cU_1)=\cU_1$.
 Eq. (\ref{defV_N}) follows from inverting the corresponding matrix.

 Suppose the Theorem to hold for $(N-1)$th   transformation steps
 meaning that the matrix $F_{N-1}$ has the entries
 \begin{equation} \label{V_N-1}
f^{N-1}_{i,j} = \frac{|W_{ij}(\cU_1, \ldots, \cU_{N-1})|}%
{|W(\cU_1, \ldots, \cU_{N-1})|}
 \end{equation}

 To prove the statement we have to calculate the value
  $\widetilde{F}_N =  Y_N' Y_N^{-1}$.
  The matrix $Y_N'$ has the derivatives (\ref{dy_jk_fin}) as the
  entries.
  Now since
   \begin{equation}
 (Y_N^{-1})_{i,j} =  \frac{1}{|Y_N|} A_{ji}
 \end{equation}
 where $A_{ij}$ is the cofactor of the element $(Y_N)_{i,j}$ in the
 matrix $Y_N$ then for the entries of the matrix $\widetilde{F}_N$
 one has
 \begin{equation}  \label{tV_N_ij}
 \begin{array}{rl}
\widetilde{f}^N_{i,j} =& \displaystyle{\frac{1}{|W(\cU_1, \ldots,
\cU_{N-1})|} \frac{1}{|Y_N|}}
\\[1em]
& \times\sum\limits_{l=1}^{n} \left[    \Gamma_{il} A_{jl} -
\sum\limits_{m=1}^n |W_{im}(\cU_1, \ldots, \cU_{N-1})| y_{m,l}
A_{jl} \right]
 \end{array}
\end{equation}
 To calculate the first term in the square brackets in (\ref{tV_N_ij}) we use the
 equation
 \begin{equation} \label{Y_N^ij}
\sum\limits_{l=1}^n \Gamma_{il} A_{jl} = |Y_N^{ij}|\,
 |W(\cU_1, \ldots, \cU_{N-1})|
\end{equation}
 where $Y_N^{ij}$ is the matrix obtained from  $Y_N$ by replacing
 in its $j$th row the detrminants $|W_j^l(\cU_1, \ldots, \cU_{N-1})|$
 (see Eq. \ref{y_jk})
 with $\Gamma_{il}$,
 $l=1,\ldots ,n$,
 which follows directly from the decomposition of the determinant
 $|Y_N^{ij}|$ on its $j$th row containing
 $\Gamma_{il}/|W(\cU_1, \ldots, \cU_{N-1})|$.
 Another relation to be used in these transformations is the
 product of a matrix with its inverse written in terms of matrix
 elements
  \begin{equation}
 \sum\limits_{k=1}^n y_{ik} A_{jk} =  \delta_{ij}
 \end{equation}
 Now we rewrite equation (\ref{tV_N_ij}) as follows
 \begin{equation} \label{tV_N_ij_fin}
\widetilde{f}^N_{ij} = \frac{|Y_N^{ij}|}{|Y_N|} -
\frac{|W_{ij}(\cU_1, \ldots, \cU_{N-1})|}{|W(\cU_1,
\ldots,\cU_{N-1})|}
\end{equation}

According to (\ref{V_N-1}) the last term in (\ref{tV_N_ij_fin})
represents the elements of the matrix $F_{N-1}$. As a result for
the entries of the matrix $F_N$ one gets
   \begin{equation}  \label{f_N}
   f^N_{i,j}= \frac{ |Y_N^{ij}|}{|Y_N|}
   \end{equation}
 The final comment is that the determinants  representing numerators of elements
 of the matrix $Y_N^{ij}$
 (we recall that it coincides with the matrix $Y_N$ composed of
 the elements (\ref{y_jk}) except for the $j$th row composed of the
 elements
 $\Gamma_{il}/|W(\cU_1, \ldots, \cU_{N-1})|$)
 are
 up to the factor $1/|W(\cU_1, \ldots, \cU_{N-1})|$ the
 minors embordering the block
  $W(\cU_1, \ldots, \cU_{N-1})$ in the matrix
  $W_{ij}(\cU_1, \ldots, \cU_N)$ and according to Sylvester
  identity one can write
  \begin{equation}
|Y_N^{ij}|  = \frac{|W_{ij}(\cU_1, \ldots, \cU_N)|}{|W(\cU_1,
\ldots, \cU_{N-1})|}. \label{Y_N^ij_a}
\end{equation}
 which together with (\ref{f_N}) and  the expression (\ref{Y_N}) for $|Y_N|$
  proves the Theorem.
 \hfill $\Box$

\section*{Acknowledgments}

The work of BFS has been partially supported by the European FEDER
and by the Spanish MCYT (Grant BFM2002-03773) and by MECD (Grant
SAB2000-0240). Authors would like to thank Fl. Stancu for helpful
comments.

 \setcounter{equation}{0}
 \renewcommand{\theequation}{A.\arabic{equation}}

 \section*{Appendix A}

 Here we formulate the Sylvester identity \cite{Gantmaher}.
 Consider a square matrix of dimension $p+q$, $p,q=1,2,\ldots $
 \be
  A=\left(
  \begin{array}{cccccc}
  a_{1,1} & \ldots & a_{1,p} & a_{1,p+1} & \ldots & a_{1,p+q}\\
         &  \ldots &       &          & \ldots &         \\
  a_{p,1} & \ldots & a_{p,p} & a_{p,p+1} & \ldots & a_{p,p+q}\\
  b_{1,1} & \ldots & b_{1,p} & b_{1,p+1} & \ldots & b_{1,p+q}\\
         &  \ldots &       &          & \ldots &         \\
  b_{q,1} & \ldots & b_{q,p} & b_{q,p+1} & \ldots & b_{q,p+q}\\
  \end{array}
  \right)
 \ee
 Let $a$ be the submatrix of dimension $p\times p$ composed of the
 elements $a_{i,j}$, $i,j=1,\ldots ,p$. If to the bottom of $a$
 we add a line of elements $b_{k,1}$, \ldots , $b_{k,p}$, to
 the right of $a$ we add a column of elements $a_{1,p+l}$, \ldots , $a_{p,p+l}$
 and the right bottom corner we fill with the element $b_{k,p+l}$,
 we obtain a square matrix $m_{j,l}$. One says that   $m_{j,l}$ is
 obtained from $A$ by embordering the block $a$ with $k$th row and
 $(p+l)$th column. The determinant $| m_{j,l} |$ is called an embordering
 minor in the determinant $|A|$. Since $k$ and $l$ can take the values
 $k,l=1,\ldots ,q$ one has $q\times q$ embordering minors from
 which one can construct the matrix $M=(m_{j,l})$. The Sylvester
 identity relates the determinants $|M|$, $|A|$ and $|a|$ as
 follows:
 \be
  |M|=|a|^{q-1}\,|A|
 \ee

\end{document}